\documentclass[twocolumn,english,reprint, longbibliography, superscriptaddress, breaklinks=true, showkeys, showpacs=false, nofootinbib]{revtex4-2}

\usepackage[T1]{fontenc}
\usepackage[utf8]{inputenc}
\usepackage{color}
\usepackage{babel}
\usepackage{amsmath}
\usepackage{amssymb}
\usepackage{subfigure}
\usepackage{graphicx}
\usepackage{physics} 
\usepackage{dsfont}
\usepackage{hyperref}

\begin{document}

\title{The irreversibility of relativistic time-dilation}

\author{Marcos L. W. Basso}
\email[Corresponding author: ]{marcoslwbasso@hotmail.com}
\affiliation{Center for Natural and Human Sciences, Federal University of ABC, Avenue of the States, Santo Andr\'e, S\~ao Paulo, 09210-580, Brazil}

\author{Jonas Maziero}
\email{jonas.maziero@ufsm.br}
\affiliation{Departament of Physics, Center for Natural and Exact Sciences, Federal University of Santa Maria, Roraima Avenue 1000, Santa Maria, Rio Grande do Sul, 97105-900, Brazil}

\author{Lucas C. C\'eleri}
\email{lucas@qpequi.com}
\affiliation{QPequi Group, Institute of Physics, Federal University of Goi\'as, Goi\^ania, Goi\'as, 74.690-900, Brazil}

\begin{abstract}
The fluctuation relations, which characterize irreversible processes in Nature, are among the most important results in non-equilibrium physics. In short, these relations say that it is exponentially unlikely for us to observe a time-reversed process and, thus, establish the thermodynamic arrow of time pointing from low to high entropy. On the other hand, fundamental physical theories are invariant under time-reversal symmetry. Although in Newtonian and quantum physics the emergence of irreversible processes, as well as fluctuation relations, is relatively well understood, many problems arise when relativity enters the game. In this work, by considering a specific class of spacetimes, we explore the question of how the time-dilation effect enters into the fluctuation relations. We conclude that a positive entropy production emerges as a consequence of both the special relativistic and the gravitational (enclosed in the equivalence principle) time-dilation effects.
\end{abstract}


\maketitle
 
\section{Introduction} 

Connections between relativity and thermodynamics can be traced back to more than one hundred years ago, when Einstein and Planck studied the question of how thermodynamic quantities, like temperature, would behave under a coordinate transformation~\cite{einstein1907,einstein1989,planck1908}. While they proposed that a moving body would appear cooler, 50 years latter exactly the opposite situation was considered by Ott~\cite{Ott1963}. The idea that temperature should be Lorentz invariant was latter reported by Landsberg~\cite{Landsberg1966}. Despite all the controversies, many developments have been achieved in the search for a relativistic description of thermodynamics in the general context, among which we mention just a few. A very interesting result in this direction is the development of the thermodynamics of black-holes~\cite{Wald1994}, a theory that was applied in Ref.~\cite{Jacobson1995} to show that Einstein's field equations can be seen as a thermodynamic equation of state, an approach latter employed in order to study the spacetime non-equilibrium properties~\cite{Eling2006}. Also, a theory of thermodynamics and statistical mechanics in the general relativistic context was considered in Refs.~\cite{Rovelli2011, Rovelli2013}. Interestingly, a statistical description of the classical gravitational field itself was proposed in Ref.~\cite{Rovelli1993}, where it was also suggested that time has a thermodynamic origin (see Ref.~\cite{Connes1994} for the quantum version of this hypothesis). We point the reader to Ref.~\cite{Padmanabhan2010} for a review on some of these topics. 

Such a problem, of connecting relativity and thermodynamics, is still open and is deeply linked with the emergence of the arrow of time. Despite the fact that the fundamental laws of nature obey the time-reverse symmetry, irreversible processes are ubiquitous in nature~\cite{zeh}. From the eggs that get cooked, but not uncooked, and the beer that sadly warms up, to the supernova and the merging of black holes, we find them all the time everywhere around us. Irreversibility, which is characterized by the positive thermodynamic entropy production~\cite{landi}, tells us that the future can be distinguished from the past. In other words, irreversibility sets the arrow of time, pointing from low to high entropy~\cite{parrondo}. 

One of the most important developments in this direction is the set of results known as fluctuation theorems, which are generalizations of the second law of thermodynamics. These theorems state that the probability of observing a negative entropy production ---i.e. the reverse of the arrow of time--- vanishes exponentially~\cite{Crooks1998,Jarzynski2011,Seifert2012,Esposito2009,Campisi2011}. One of the implications of this result is that on average a positive entropy production will always be observed for any process.

Here we are interested in how relativity affects such relations. Some progress in this direction have been achieved. The entropy production of a spin system moving in a gravitational field was considered in Ref.~\cite{Esfahani2007}. The entropy, in this case, emerges as a consequence of the entanglement between the spin and the momentum degrees of freedom. Considering the linear regime, a fluctuation-dissipation relation for gravity was derived in Ref.~\cite{Mottola1986}. The non-equilibrium fluctuations of a black hole horizon were investigated in Ref.~\cite{Iso2011} by means of the Jarzynski equality~\cite{Jarzynski1997} in connection with the generalized second law of thermodynamics~\cite{Bekenstein1974}. 

Specifically, the main goal of the present work is to consider the question if entropy can be produced in a closed quantum system that lives in a spacetime where time-dilation is allowed. We consider a quantum system that moves in a spacetime under the influence of the special relativistic and the gravitational time-dilation and show that, in general, irreversibility will arise. Such result, which cannot be explained without quantum mechanics, special relativity and the equivalence principle, raises several other questions that we discuss at the end of this article.

In the next section, we first present a derivation of a general fluctuation theorem in the usual form, without considering time-dilation. After this, the corrections coming from time-dilation are included and we present our main result. A physical discussion, including some open questions, is left to the last section of the paper. We use natural units in which the speed of light $c$, Planck's constant $\hbar$, Newton's gravitational constant $G$ and Boltzmann constant $k_{B}$ are set to unity. The signature of the metric is $(-,+,+,+)$. 

\section{Entropy production}

We start this section by presenting the derivation of Jarzynski's equality on flat spacetimes, without considering any effect of time-dilation. Although well known, this is done here for establishing the setup we are considering as well as to fix the notation. After this, based on the same setup, we present the derivation of the fluctuation relation when both the gravitational and the special relativistic time-dilation enter the game.

\subsection{The fluctuation theorem}
\label{sec:IIa}
Let us consider the following protocol. First, at the initial time $t=0$, a quantum system is prepared in the thermal state 
\begin{equation}
\omega_{0} = \frac{1}{Z_{0}}e^{-\beta H_{0}},
\end{equation}
with $H_{0}$ and $Z_{0} = \Tr{e^{-\beta H_{0}}}$ being the initial Hamiltonian and partition function, respectively. $\beta$ is the inverse temperature, which is usually associated with some reference environment. Since the Hamiltonian is Hermitian, we can write its spectral decomposition as $H_{0}\ket{\epsilon_{m}^{0}} = \epsilon_{m}^{0}\ket{\epsilon_{m}^{0}}$, and thus the energy projection operator are defined by $\Pi_{m}^{0} = \ket{\epsilon_{m}^{0}}\bra{\epsilon_{m}^{0}}$. The second step of the protocol consists in a measurement in this basis, resulting in the energy eigenvalue $\epsilon_{m}^{0}$ with probability
\begin{equation}
p_{m} = \frac{e^{-\beta\epsilon_{m}^{0}}}{Z_{0}}.
\end{equation}

The system, which is now in the pure state $\ket{\epsilon_{m}^{0}}$, undergoes a time evolution ---during the period $T$--- governed by the completely-positive and trace-preserving map $\Theta$, which can be written in terms of the Kraus's operators $K_{j}$ as $\Theta(\rho) = \sum_{j} K_{j}\rho K_{j}^{\dagger}$, with $\rho$ being the initial state of the system~\cite{Nielsen2010}. Preservation of probability demands that $\sum_{j} K_{j}^{\dagger}K_{j} = \mathds{1}$. If $\sum_{j} K_{j}K_{j}^{\dagger} = \mathds{1}$ also holds, the map is said to be unital and the identity is preserved, i.e.,  $\Theta(\mathds{1}) = \mathds{1}$. During this process, the Hamiltonian changes from $H_{0}$ to $H_{T}$, and some work is performed. 

In the final step, at $t=T$, the system is measured in the eigenbasis of the final Hamiltonian, resulting in the eigenvalue $\epsilon_{n}^{T}$ with conditional probability
\begin{equation}
    p_{n|m}^{T} = \Tr{\Pi_{n}^{T}\Theta\left[ \Pi_{m}^{0}\right]},
\end{equation}
where $\Pi_{n}^{T} = \ket{\epsilon_{n}^{T}}\bra{\epsilon_{n}^{T}}$ is the final energy projector defined in terms of the final Hamiltonian eigenbasis: $H_{T}\ket{\epsilon_{n}^{T}} = \epsilon_{n}^{T}\ket{\epsilon_{n}^{T}}$. From this, we can construct the joint probability of obtaining $\epsilon_{m}^{0}$ in the first measurement and $\epsilon_{n}^{T}$ in the second one as $p_{m,n} = p_{m}p_{n|m}^{T}$. 

Now, by defining the work as the stochastic variable $W_{n,m} = \epsilon_{n}^{T} - \epsilon_{m}^{0}$, the work probability distribution density can be defined as
\begin{equation}
    P(W) = \sum_{m,n}p_{m,n}\delta\left[W - W_{n,m}\right].
\end{equation}
From these definitions we obtain the generalized Jarzynski equality~\cite{Rastegin2014}
\begin{eqnarray}
\expval{e^{-\beta W}} &=& \int \dd W e^{-\beta W} \nonumber \\
&=&e^{-\beta\Delta F}\left[1 + \Tr{G_{\Theta}\omega_{T}}\right],
\label{eq:jar}
\end{eqnarray}
where $G_{\Theta} = \Theta(\rho_{*}) - \rho_{*}$, with $\rho_{*}$ being the maximally mixed state, quantifies how much the map $\Theta$ deviates from unitality. Besides, $\omega_{T} = Z_{T}^{-1}e^{-\beta H_{T}}$ is the reference final thermal state, while the free energy difference is defined as $\Delta F = F_{T} - F_{0}$ with $F_{x} = -\beta^{-1}\ln{Z_{x}}$. If the map is unital, then $G_{\Theta} = 0$ and we recover the original Jarzynski equality~\cite{Jarzynski1997}. 

Therefore, considering unital processes, thermodynamic entropy will be produced,
\begin{equation}
    \expval{\Sigma} = \beta\left(\expval{W} - \Delta F\right) \geq 0,
    \label{eq:second_flat}
\end{equation}
as long as the process takes the system out of equilibrium. This equation, which is a statement of the second law of thermodynamics, can be seen to be a consequence of Eq.~\eqref{eq:jar} (with $G_{\Theta} = 0$) by taking into account the convexity of the exponential function, which allows us to employ Jensen's inequality.

The goal of the next sub-section is to consider the corrections to Eqs.~\eqref{eq:jar} and~\eqref{eq:second_flat} introduced by the gravitational and the special relativistic time-dilation.

\subsection{Fluctuation theorem and time-dilation}
\label{sec:IIb}

In this section, we consider the Hamiltonian formulation of the dynamics of a quantum particle with some internal structure in a curved spacetime. In order to do this, we follow Refs.~\cite{Zych2011,Pikovski2015} and begin by considering the classical description, which we latter quantize. 

Let $(\mathcal{M}, \mathbf{g})$ be a time-orientable spacetime~\cite{Wald}, with $\mathcal{M}$ being a differential manifold and $\mathbf{g}$ a Lorentzian metric. Here, we demand that $\mathcal{M}$ to be time orientable since this is necessary for defining our thermal equilibrium states. As will be specified later, we restrict ourselves to a static spacetime and then consider the Newtonian limit for the derivation of our main result. Since we will be dealing with a static spacetime, in which there exists a time-like Killing vector field $\xi$ orthogonal to the simultaneous hyper-surfaces $t = \text{constant}$ \cite{Wald}, there will be a preferential notion of time orientability, that we use in order to define local thermal equilibrium states \cite{Rovelli2011, Rovelli2013}. Let us then consider a particle with some internal structure (internal degrees of freedom) traveling along the world-line $\gamma : I\subset \mathbb{R} \to \mathcal{M}$ on this spacetime. Note that this is a semiclassical system in the sense that while its external degrees of freedom have a well defined trajectory in spacetime, its internal degrees of freedom will be described quantum mechanically. 

The idea here is to pick up a coordinate system $x^{\mu}$ ($\mu$ running from 0 to 3 and we denote $x^{0}\equiv t$), with respect to which we describe the $4$-momentum of the particle along the world-line $\gamma$ as $p^{\mu}$. Such coordinate system contains the symmetries of the spacetime, including the time-like Killing field $\xi$ that will be employed in order to define the static observers at each point of the spacetime, named the laboratory frames. In these frames we can express the Killing vector field as $\xi = \partial_t$, or equivalently, $\xi^{\mu} = \delta^{\mu}_{\ t}$. Therefore, the static observers can be defined by the normalized $4$-velocity $u^{\mu} = \xi^{\mu}/ \sqrt{- \xi_{\mu}\xi^{\mu}}$. All these laboratory frames has the same time-orientability and therefore we can define a thermal equilibrium state that will be used as the reference state. Now, in order to define the Hamiltonian of our system, we move to a comoving reference frame, which we denote by primed coordinates $x^{\mu'}$. Such change is necessary for us to split the dynamics of the internal and the external degrees of freedom (the centre-of-mass) of our system. After this, we come back to the coordinate system $x^{\mu}$ that describe our system from the laboratory point of view. 

In the rest-frame of the particle (the point of view of the comoving observer), we have that $p^{j'} =  (\partial x^{j'}/\partial x^{\mu}) p^{\mu} = 0$ (with $j$ labeling only the spatial coordinates). Then, the total energy as measured by the comoving observer is given by $p_{t'}$ ($x^{0'} \equiv t'$). It comprises not only the energy stemming from the rest mass of the system but also any binding or kinetic energies of the internal degrees of freedom and thus also the particle's internal Hamiltonian $H_{\text{int}}$. Therefore
\begin{align}
    p_{t'} = m + H_{\text{int}} \equiv H_{\text{rest}}. \label{eq:Hrest}
\end{align}
On the other hand, $p^t$ describes the dynamics of the particle with respect to the observer associated with the coordinate system $x^{\mu}$, which includes the energy of both internal and external
degrees of freedom. Therefore, it constitutes the total Hamiltonian of the system relative to $x^{\mu}$ and will be denoted by $H = p_t$. Given that $p_{\mu} p^{\mu} = p_{\mu'}p^{\mu'}$ is a coordinate invariant quantity, we have the following relation between $H$ and $H_{\text{rest}}$
\begin{align}
    H = \sqrt{\frac{g^{t't'}H^2_{\text{rest}} - p_j p^j}{g^{tt}}}.
\end{align}
Taking the component $x^{t'}$ associated with the comoving observer to be the proper time $\tau$ along the particle's world line implies that $g^{t't'} = \eta^{tt} = -1$, and therefore
\begin{align}
    H = \sqrt{-\frac{H^2_{\text{rest}} + p_j p^j}{g^{tt}}}.
\end{align}
Note that the equations in this paragraph are valid for any spacetime.

Now we restrict ourselves to a static spacetime, for which there is a time-like Killing vector field orthogonal to the hyper-surfaces $t = \text{constant}$, which implies that the metric can be written as $\mathbf{g} = g_{tt} \dd t \otimes \dd t + h_{ij} \dd x^i \otimes \dd x^j$, with $\mathbf{h}$ being the induced Riemannian metric on the hyper-surfaces $t = \text{constant}$~\cite{Wald}. For this kind of spacetime, $(g_{tt})^{-1} = g^{tt}$, and we can write
\begin{align}
    H = \sqrt{-g_{tt}(H^2_{\text{rest}} + p_j p^j) }.
\end{align}
For instance, in the Newtonian limit, which is characterized by a weak and static gravitation field, and considering that the system moves slowly when compared to the speed of light, we obtain $g_{tt} \approx - (1 + 2 \phi)$ and $g_{ij} \approx \delta_{ij}$, where $\phi$ is the usual Newtonian gravitational potential. As a consequence, the total Hamiltonian can be expressed as
\begin{equation}
    H \approx H_{\text{rest}} + \frac{p^2}{2 H_{\text{rest}}} + H_{\text{rest}}\phi(x). 
\end{equation}
Now, by using Eq.~\eqref{eq:Hrest} at the lowest order we obtain
\begin{equation}
    H = H_{\text{cm}} + H_{\text{int}}\Big(1 + \phi(x) - \frac{p^2 }{2m^2}\Big) \label{eq:hamil_},
\end{equation}
with $H_{\text{cm}} = m + p^2/2m + m \phi(x)$. Moreover, by taking into account that $\dd\tau^2 = -g_{\mu \nu} \dd x^{\mu} \dd x^{\nu}$, it is interesting to observe that
\begin{align}
    \dv{\tau}{t} & = \sqrt{-g_{tt} - g_{ij}\dv{x^i}{t}\dv{x^j}{t}}\nonumber\\
    & \approx 1 + \phi(x) - \frac{p^2 }{2m^2},
\end{align}
which implies that Eq.~(\ref{eq:hamil_}) can be recast as
\begin{align}
H  = H_{\text{cm}} + H_{\text{int}}\dv{\tau}{t}. \label{eq:hamil1}
\end{align}
This equation implies that the internal Hamiltonian is effectively re-scaled by the factor $\dd\tau/\dd t$ when the internal evolution is described with respect to the coordinate time $t$ regarding the laboratory frame (the coordinates $x^{\mu}$).

If the particle under consideration is a quantum system with some internal degree of freedom, e.g., a spin, we can quantize this classical scheme. Let us consider a general quantum system characterized by the Hilbert space $\mathcal{H_{\text{cm}}} \otimes \mathcal{H}_{\text{int}}$ consisting of the composite Hilbert space of the centre-of-mass and the internal degrees of freedom. The total Hamiltonian $H$, given in Eq.~\eqref{eq:hamil1}, is promoted to a Hermitian operator that acts in this composite Hilbert space, with the operator $H_{\text{cm}}$ describing the dynamics of the external degrees of freedom of the particle while the second term in the right-hand side of Eq.~\eqref{eq:hamil1} describes the dynamics of the internal degrees of freedom.

If we start with an uncorrelated state, the initial state of the quantum system at the point $\mathfrak{p} \in \mathcal{M}$ is given by
\begin{align}
    \ket{\Psi_0} = \ket{x_0}\otimes \ket{\psi_0}.
\end{align}
This is not an actual restriction since the preparation of the initial state usually requires some measurement that destroys all the correlations between the internal and external degrees of freedom.

Now, as the quantum system travels along its world line $\gamma$, the evolved state will be given by
\begin{align}
     \ket{\Psi(t)} &= e^{-\frac{i}{\hbar}\int_{\gamma} H dt} \ket{x_0}\otimes \ket{\psi_0} \nonumber \\
     & = e^{-i\int_{\gamma} H_{\text{cm}} + H_{\text{int}}\Big(1 + \phi(x) - \frac{p^2 }{2m^2}\Big) dt} \ket{x_0}\otimes \ket{\psi_0} \nonumber \\
     & = e^{-i\int_{\gamma} H_{\text{cm}} dt} \ket{x_0} \otimes e^{-i\int_{\gamma}  H_{\text{int}}\frac{d \tau}{dt} dt} \ket{\psi_0}.
\end{align}
The last equality follows from the semiclassical approximation, in which the motion of the quantum particle along its world-line is well-defined. Under this approximation, the terms in $H = H_{\text{cm}} + H_{\text{int}} (\dd \tau/\dd t)$, apart from $H_{\text{int}}$, are fixed functions along the path $\gamma$. Therefore, we can see that the evolution of the internal degrees of freedom is unitary and given by
\begin{align}
    \ket{\psi_{\tau}} & = e^{-i\int_{\gamma}  H_{\text{int}}\Big(1 + \phi(x) - \frac{p^2 }{2m^2}\Big)  dt} \ket{\psi_0}\nonumber \\ 
    & = e^{-i\int_{\gamma}  H_{\text{int}} d\tau} \ket{\psi_0},
    \label{eq:evolution}
\end{align}
i.e., the internal degrees of freedom evolve with respect to the proper time and along the system's world line $\gamma$. Regarding the laboratory frame, one can see that the internal Hamiltonian is effectively rescaled by the factor $\dd\tau/\dd t$ when the internal evolution is described with respect to the coordinate time $t$. 

Going back to the work protocol discussed in the previous section, given that $H_0 = H_{\text{int}}(\dd\tau/\dd t) = H_{\text{int}}$ is the initial Hamiltonian at the point $\mathfrak{p} \in \mathcal{M}$, one define $H_{\text{int}}\ket{\epsilon_{m}^{0}} = \epsilon_{m}^{0}\ket{\epsilon_{m}^{0}}$ by choosing $\phi(\mathfrak{p}) \equiv 0$. Since the particle is initially at rest, we have $\dd\tau/\dd t  = 1$. At some latter proper time, when the system is found at the point $\mathfrak{q} = \gamma(\tau) \in \mathcal{M}$ of the trajectory, the Hamiltonian is given by $H_{\tau} = H_{\text{int}}(\dd \tau/\dd t)$, with  $H_{\tau}\ket{\epsilon_{m}^{\tau}} = \epsilon_{m}^{\tau}\ket{\epsilon_{m}^{\tau}}$ where
\begin{align}
    \ket{\epsilon_{m}^{\tau}} = e^{-i\int_{\gamma}  H_{\text{int}} \dd\tau}\ket{\epsilon_{m}^{0}}. \label{eq:eigenevo}
\end{align}
By assuming that $H_{\text{int}} $ is time-independent, we can see that the eigenvalues of the Hamiltonian are rescaled by the factor $\dd \tau/\dd t$:
\begin{align}
H_{\tau}\ket{\epsilon_{m}^{\tau}} & = H_{\text{int}}\dv{\tau}{t}  e^{-i\int_{\gamma}  H_{\text{int}} \dd\tau}\ket{\epsilon_{m}^{0}}  \nonumber\\
    & =  \dv{\tau}{t} \epsilon_{m}^{0} \ket{\epsilon_{m}^{\tau}}.
    \label{eq:hamil}
\end{align}
Therefore
\begin{align}
     \epsilon_{m}^{\tau} = \dv{\tau}{t} \epsilon_{m}^{0} \approx \Big(1 + \phi(x) - \frac{p^2 }{2m^2}\Big)\epsilon_{m}^{0}.
     \label{eq:eigenvalue}
\end{align}
The case of time-dependent Hamiltonian will be addressed in the appendix.

Now we are ready to describe the work protocol, discussed earlier, in this context. The system starts in the thermal state with some inverse temperature $\beta$ at $\tau = 0$\footnote{Since we are working with a static spacetime and under the Newtonian approximation, the fact that we need a preferred notion of time in order to define the thermal state is not important here. See Refs.~\cite{Rovelli1993,Connes1994} for more details on this problem.}. This is well defined from the point of view of the laboratory observers \cite{Rovelli2011, Rovelli2013}. After the preparation, a projective energy measurement (defined by $H_{\text{int}}$) is performed on the system and we assume that such action takes no time. If the eigenvalue $\epsilon_{m}^{0}$ is obtained, the state of the system just after the measurement is given by $\ket{\epsilon_{m}^{0}}$. The external degrees of freedom do not enter here since we are considering the semiclassical approximation. We then let the system to evolve by following the trajectory $\gamma$ on spacetime. This is our process $\Theta$, whose dynamics is unitary and given by Eq.~\eqref{eq:evolution}. After a certain amount of proper time $\tau$, the second energy measurement is performed (according to the final Hamiltonian $H_{\tau}$), resulting in the eigenvalue $\epsilon_{n}^{\tau}$. From Eqs.~\eqref{eq:hamil} and~\eqref{eq:eigenvalue} we can define the work as
\begin{equation}
W_{n,m} = \epsilon_{n}^{\tau} - \epsilon_{m}^{0} =  \left(\dv{\tau}{t}\epsilon_{n}^{0} - \epsilon_{m}^{0}\right)\delta_{n,m}. 
\label{eq:work}
\end{equation}

Now, given the class of spacetimes considered here, the initial probability ---of obtaining the eigenvalue $\epsilon^{0}_{m}$--- is well defined by the thermal distribution. The conditional probability of obtaining the eigenvalue $\epsilon^{\tau}_{n}$ can be computed by considering the evolution of $\ket{\epsilon_{m}^{0}}$ under the dynamics given in Eq.~\eqref{eq:evolution}. From these probabilities, and defining $\Delta F_{\gamma} = F_{\tau} - F_{0}$, we can build the joint probability of both measurements and write down the fluctuation relation as
\begin{equation}
    \expval{e^{-\beta W}}_{\gamma} = e^{-\beta\Delta F_{\gamma}},
    \label{eq:jar_time}
\end{equation}
with $F_{\tau} = -\beta^{-1}\ln Z_{\tau}$ carrying the contribution coming from time-dilation. The $\gamma$ subscript in the average above remembers us that the joint probability distribution depends on the path the system follows on the spacetime, as can be seem from Eq.~\eqref{eq:evolution}. Another observation is that the final temperature we consider here is the same as the initial one since this is just a reference state, defined at the beginning of the process by the laboratory observers. For instance, we can consider that the initial and final points where the projective measurement are realized lay in the world-line of just one of the laboratory observers or that the initial and final points of the world-line of the system lay in different laboratory observers. So, we are asking by how much the system was took out from this reference equilibrium state due to time-dilation. The answer to this question is given by Eq.~\eqref{eq:jar_time} and shows the fact that gravity couples the internal degrees of freedom of the system with the path, which means that some information about the system must be retained in the field (the underlying spacetime), i.e., information has been exchanged. The resulting effect is the shift of the energy eigenvalues, modifying the entropy of the system. 

It is important to observe here that, depending on the sign of $\dd \tau/\dd t$, the free energy change $\Delta F_{\gamma}$ can be negative, positive or zero. If we consider the system is comoving with the static observers, we obtain $\dd \tau/\dd t = 1$ and, consequently, $\Delta F_{\gamma} = 0$. For this case we do not expect any entropy being produced since there will be no measurable time-dilation. If $\dd \tau/\dd t > 1$, then $\Delta F_{\gamma}$ will be positive and Eq.~\eqref{eq:jar_time}, along with Jensen's inequality, leads us to
\begin{equation}
\expval{W}_{\gamma} \geq \Delta F_{\gamma},
\label{eq:second}
\end{equation}
which is the second law of thermodynamics enclosing time-dilation. This is the case, for instance, of the system moving to a region of spacetime where its energy eigenvalues get blue-shifted, i.e., $\epsilon^{\tau}_m > \epsilon^{0}_m$. 

The last possibility happens for trajectories for which $\dd \tau/\dd t < 1$, implying that $\Delta F_{\gamma} < 0$. However, Eq.~\eqref{eq:work} tells us that the work also changes sign. The difference here is that the work is being performed by the system against the field. By taking into account Jensen's inequality, we obtain $\expval{W}_{\gamma} \geq \Delta F_{\gamma}$. Note that, in this case, both $\expval{W}_\gamma$ and $\Delta F_{\gamma}$ are negative quantities. But the energy flux can be defined to be positive for any of the directions (from system to the field or from the field to the system). Therefore, we obtain the same result when considering the red-shift case. This is analogous to the well-known result that a photon has to expend energy to climb the gravitational (metric) potential~\cite{Wald}. 

In order to illustrate this point, let us consider the simple example where the internal degree of freedom of the system is a harmonic oscillator whose eigenvalues are given by $\epsilon_n^{0} = (n + 1/2) \omega$, with integer $n\geq 0$ while $\omega$ stands for the initial frequency of the oscillator as measured by the laboratory observer. A direct calculation shows that
\begin{align}
    \beta\Delta F_{\gamma}  = \ln \left[\frac{\sinh{(\frac{d \tau}{dt} \frac{\beta  \omega}{2})}}{\sinh{(\frac{\beta \omega}{2})}} \right],
\end{align}
where we can directly identify the three cases discussed above. For instance, let us consider a scheme regarding only the gravitational time dilation. To do this we can consider the expansion given in Eq.~\eqref{eq:hamil_} in the limit of large $m$, for which the kinematic degrees of freedom can be ignored and therefore $d \tau / d t  \approx \sqrt{-g_{tt}} \approx 1 + \phi(x)$. From the point of view of the Killing static observers, $\Delta F_{\gamma} > 0 $ for the trajectories that $\abs{g_{tt}}$ increases along the world line of the system, while $\Delta F_{\gamma} < 0 $ for the trajectories that $\abs{g_{tt}}$ decreases along the world line of the system. As well, if the system is comoving with the static observers, there is no variation of the gravitational potential along the trajectory of the system, which implies that $\dd\tau /dt = 1$ and thus $\Delta F_{\gamma} = 0$. On the other hand, a direct calculation also shows that $\expval{W}_{\gamma} = (\dd \tau/\dd t - 1)\epsilon^{0}_n$, which has the same interpretation as $\Delta F_{\gamma}$. Finally, going back to the work protocol, we considered that the first measurement in the energy eigenbasis is performed at point $\mathfrak{p} \in \mathcal{M}$ where $\phi(\mathfrak{p}) \equiv 0$, while the second measurement in occurs at point $\mathfrak{q} \in \mathcal{M}$. Therefore, more generally, we can write $\expval{W}_{\gamma}/\epsilon^{0}_n = (\dd \tau/\dd t)|_{\mathfrak{q}} - (\dd \tau/\dd t)|_{\mathfrak{p}} = \phi(\mathfrak{q}) - \phi(\mathfrak{p})$,  which implies that the change of the gravitional potential is codified in the change of the locally measured energy of the internal degree of freedom of the system.

\section{Discussion}

We considered the problem of entropy production of a semiclassical particle ---whose external degrees of freedom are described by classical mechanics while the internal ones follow the laws of quantum mechanics--- travelling in a static spacetime under Newtonian approximation. We derived an integral fluctuation relation by considering the particle dynamics on this spacetime as the work process. The result, shown in Eq.~\eqref{eq:jar_time}, encompass the corrections due to time-dilation coming from both the equivalence principle and the special theory of relativity. As a check of consistency, if we take the limit in which the speed of light goes to infinity,  $c\rightarrow\infty$, thus entering in the Newtonian physics, both effects vanish, and we recover the trivial result $\expval{W}_{\gamma} = \expval{W} \geq 0$, independently of the trajectory of the particles. This is expected since we do not have the effect of time-dilation in Newtonian physics. The same result is obtained when we consider that the particle is static and thus comoving with the Killing static observers.

However, if we consider any other trajectory there will be entropy being produced due to time dilation. This happens because the system couples to the path and, thus, some information about the system must be retained in the field (the underlying spacetime), thus modifying the entropy of the system.

Although we consider in the main text an autonomous quantum system (time-independent Hamiltonian), the case of a driven system, for which the Hamiltonian depends on the proper time, is presented in the Appendix. Qualitatively, the result is the same, but now we have a contribution of the driven part of the Hamiltonian, thus modifying the entropy production rate.

A natural question that arises is the extension of our result to a general spacetime, for which we expect that the right-hand-side of Eq.~\eqref{eq:jar_time} will depend on the spacetime curvature, thus taking into account the full action of the gravitational field, going beyond the equivalence principle presented here.

Another issue is the semiclassical approximation. If we consider that the external degrees of freedom of the system are described by quantum mechanics, superpositions of distinct trajectories will be allowed, bringing new phenomena to light~\cite{Zych2011,Roura2020}. One interesting question in this direction is the role of the entanglement between the internal and external degrees of freedom. Also, we can explore the role of quantum coherences in such scenarios~\cite{Santos2019}, specially in the context of ergotropy~\cite{Francica2020}. 

In conclusion, thermodynamic entropy will in general be produced in a quantum system due to time-dilation, which is a consequence of the causal structure of the spacetime introduced by Einstein. As we only consider the time dilation, gravity, as understood in Einstein general theory of relativity, is not included in our analysis. However, our results cannot be explained without effects coming from quantum mechanics, special relativity and the equivalence principle. This is a very fundamental effect that says that a local observer will perceive a positive entropy production in the evolution of a closed system just because he lives in a Lorentzian spacetime, where the speed of light is finite. 

This is a very interesting conclusion, that says that the thermodynamic arrow (from low to high entropy) is rooted into the causal structure of spacetime and, as a consequence, also is the so called arrow of causation (from effect to cause), which is based on the thermodynamic one. In Ref.~\cite{Rovelli2022}, the author provides a deeper discussion on the connection between these arrows, while here we link both to time-dilation.

\begin{acknowledgments}
This work was supported by the S\~{a}o Paulo Research Foundation (FAPESP), Grant No.~2022/09496-8, by the National Institute for the Science and Technology of Quantum Information (INCT-IQ), Grant No.~465469/2014-0, and by the National Council for Scientific and Technological Development (CNPq), Grants No.~309862/2021-3 and No~308065/2022-0.
\end{acknowledgments}

\appendix*
\section{Proper time-dependent Hamiltonian}
\label{app:time}

In this appendix, for completeness, we discuss the derivation of Jarzynski equality under time dilation for the general case in which the Hamiltonian of the internal degrees of freedom depends on the proper time, i.e., $H_{\text{int}} = H_{\text{int}}(\tau)$ such that $H_{\tau} = H_{\text{int}}(\tau)(\dd \tau/\dd t)$ is the Hamiltonian as described by the static observers. 

As in Sec.~\ref{sec:IIb}, at the point $\mathfrak{p} \in \mathcal{M}$ we can define the initial Hamiltonian as $H_0 = H_{\text{int}}(\dd\tau/\dd t) = H_{\text{int}}$. A projective measurement on the eigenbasis of $H_{\text{int}}$ is performed on the system. If the eigenvalue $\epsilon_{m}^{0}$ is obtained, the state of the system just after the measurement is given by $\ket{\epsilon_{m}^{0}}$. However, we now consider that the evolution of the internal degrees of freedom is governed by the Hamiltonian $H_{\tau} = H_{\text{int}}(\tau)(\dd \tau/\dd t)$ as described by the static observers. At some latter proper time, when the system is found at the point $\mathfrak{q} = \gamma(\tau) \in \mathcal{M}$ of the trajectory, a measurement on the eigenbasis of $H_{\tau}$ is performed, with the state vectors given by Eq.~\eqref{eq:eigenevo}. Therefore, the conditional probability of finding the eigenvalue $\epsilon_{n}^{\tau}$ is given by
\begin{align}
     p_{n|m}^{\tau} & = \abs{\braket{\epsilon_n^{\tau}} {\epsilon_m^{0}}}^2 \nonumber \\& = \abs{\bra{\epsilon_m^{0}} \mathcal{T} e^{-\frac{i}{\hbar}\int_{\gamma}  H_{\text{int}}(\tau) \dd\tau}\ket{\epsilon_n^{0}}}^2,
\end{align}
where $\mathcal{T}$ is the time ordering operator. From this, we can construct the joint probability of obtaining $\epsilon_{m}^{0}$ in the first measurement and $\epsilon_{n}^{\tau}$ in the second one as $p_{m,n} = p_{m}p_{n|m}^{\tau}$ and define the work as the stochastic variable $W_{n,m} = \epsilon_{n}^{\tau} - \epsilon_{m}^{0}$. The rest of the derivation of the Jarzynski equality follows exactly as discussed in Sec.~\ref{sec:IIa}. The only difference here is that the relation between $\epsilon_{n}^{\tau}$ and $\epsilon_{m}^{0}$ given by Eq.~\eqref{eq:eigenvalue} is no longer valid.

Finally, it is worth mentioning that if the world-line $\gamma$ of the system coincides with the world-line of one of the laboratory observers described by the four-velocity $u^{\mu} = \xi^{\mu}/ \sqrt{- \xi_{\mu}\xi^{\mu}}$, then $d \tau/ dt = 1$ and $H_{\tau} = H_{\text{int}}(\tau)$. Therefore, the protocol follows just as in (locally) flat spacetime with no effect due to relativistic time dilation. In contrast, if the world-line $\gamma$ of the system does not coincide in general with the world-line of one of the laboratory observers, but there is an intersection between both trajectories at the points $\mathfrak{p}$ and $\mathfrak{q}$ where the initial and final projective measurements in the eigenbasis of the Hamiltonian are performed, then $H_{\tau} = H_{\text{int}}(\tau)(d \tau/ dt)$ and the protocol follows exactly as described in this section. Therefore, one can compare both protocols to unveil the effect of relativistic time dilation on entropy production.


\end{document}